\documentclass[conference]{IEEEtran}

\usepackage{graphicx}
\usepackage{cite}

\usepackage{amsmath}
\usepackage{amsfonts}
\usepackage{amssymb}
\usepackage{subfigure}
\usepackage{psfrag}
\usepackage{color}
\usepackage{tikz}

\usepackage[english]{babel}
\usepackage[utf8]{inputenc}

\newcommand{\bm}{\mathbf}
\newcommand{\be}{\begin{equation}}
\newcommand{\ee}{\end{equation}}
\newcommand{\bea}{\begin{eqnarray}}
\newcommand{\eea}{\end{eqnarray}}
\newcommand{\x}{{\bm x}}

\newcommand{\bI}{{\bm I}}

\newcommand{\bF}{{\bf F}}

\newcommand{\n}{{\bm n}}

\begin{document}

\title{Blind Channel Estimation for Massive MIMO: A Deep Learning Assisted Approach}

\author{\normalsize Parna Sabeti$^{\star}$, Arman Farhang$^{\dagger}$,Irene Macaluso$^{\star}$, Nicola Marchetti$^{\star}$ and Linda Doyle$^{\star}$ 
\\$^{\star}$ CONNECT, Trinity College Dublin, Ireland \\
$^{\dagger}$ CONNECT, Maynooth University, Ireland \\
Email: sabetip@tcd.ie,
arman.farhang@mu.ie,
 \{macalusoi, nicola.marchetti,  ledoyle\}@tcd.ie\vspace{-4.5 mm}}

\maketitle
\begin{abstract}
Large scale multiple-input multiple-output (MIMO) or Massive MIMO is one of the pivotal technologies for future wireless networks. However, the performance of massive MIMO systems heavily relies on accurate channel estimation. While the acquisition of channel state information (CSI) in such systems requires an increasingly large amount of training overhead as the number of users grows. To tackle this issue, in this paper, we propose a deep learning assisted blind channel estimation technique for orthogonal frequency division multiplexing (OFDM) based massive MIMO systems. We prove that by exploiting the asymptotic orthogonality of the massive MIMO channels, the channel distortion can be averaged out without the prior knowledge of channel impulse responses, and after some mathematical manipulation, different users’ transmitted
data symbols can be extracted. Thus, by deploying a denoising convolutional neural network algorithm (DnCNN), we mitigate a remaining channel and noise effect to accurately detect the transmitted data symbols at the channel sounding stage. Using the detected data symbols as virtual pilots, we estimate the CSI
of all the users at each BS antennas. Our simulation results testify the efficacy of our proposed technique and demonstrate that it can provide a mean square error (MSE) performance which coincides with that of the data-aided channel estimation
technique. 

\end{abstract}
 \section{Introduction}\label{sec:Intro}

Large scale multiple-input multiple-output (MIMO) or massive MIMO
is one of the pivotal technologies for future wireless networks \cite{andrews2014will,larsson2014massive,series2015imt,wang2014cellular}. 
Massive MIMO can tremendously increase spectral efficiency by enabling users to simultaneously utilize the entire available bandwidth \cite{rusek2013scaling,marzetta2010noncooperative, bjornson2018massive}. Moreover, by increasing the number of base station (BS) antennas, i.e., array gain, the required power to achieve the desired information rate is reduced and energy efficiency is enhanced compared to small scale antenna systems \cite{ngo2013energy}.
Large-scale antenna arrays also play a vital role in achieving
highly directional beamforming gains in millimeter wave (mmWave) systems \cite{andrews2017modeling}.
All the aforementioned benefits of massive MIMO relies on accurate channel estimates for detection and beamforming. This is because massive MIMO is using the channel gains between users and the BS antennas to separate different users signals from one another. 
As the scale of the antenna array increases and/or the number of users in the network grows, the 
channel estimation
becomes a bottleneck for massive MIMO systems \cite{marzetta2006much}. 

Many studies have been conducted to design efficient and reliable techniques for channel estimation of massive MIMO systems. 
There are two class of channel estimation; 
data-aided and blind estimation techniques.
%

In data-aided techniques, a substantial portion of resources is reserved for transmission of known pilot signals \cite{marzetta2006much, yuan2018fundamental,bjornson2015massive}. In \cite{bjornson2015massive}, it is shown that up to half of the coherence time may be dedicated to pilot transmission.  

%
%
Blind channel estimation techniques are proposed to reduce the pilot overhead for channel acquisition.
In \cite{mezghani2017blind}, the authors 
exploit massive MIMO channel sparsity in the angular domain to blindly estimate the channels. 
However, the proposed technique in \cite{mezghani2017blind} is an iterative solution, and singular value decomposition of the covariance matrix of the received signal should be calculated in each iteration which is a computationally complex process. In \cite{zhang2017blind}, an alternative blind estimation technique is proposed that highly depends on the sparsity level of the channels. Moreover, since sparse matrix factorization is a non-convex problem, the authors proposed a projection-based bilinear generalized approximate message passing (P-BiG-AMP) algorithm as an approximate solution which imposes a considerable amount of computational complexity to the system. 


Recently, the advent of open-source deep learning (DL) libraries
has motivated another line of research for channel estimation \cite{raina2009large}. 
In \cite{ye2017power} and \cite{soltani2019deep} DL-based channel estimation techniques are proposed for single antenna systems which are not suitable for large scale antenna systems.
In \cite{he2018deep}, learned denoising-based approximate message passing (LDAMP) neural network, which is used for compressive image recovery, is applied to estimate the channel in an iterative manner for mm-Wave massive MIMO systems. Besides its high computational complexity, this work has studied only a single user scenario. 
In \cite{neumann2018learning}, a low-complexity channel estimator with an embedded neural network is proposed for massive MIMO systems. This solution assumes flat fading channel in a single user scenario, and it still requires pilot signals to estimate the channel. 
In \cite{huang2018deep}, a framework that integrates DL into massive MIMO is proposed to estimate direction of arrival (DOA) and channel impulse responses between the users and BS antennas. In fact, a deep neural network (DNN) replaces the whole massive MIMO receiver and is trained to estimate DOA. Then, the complex gains of the channels are calculated using the estimated DOA and a short training sequence. Moreover, if the number of BS antennas increases, the size of DNN layers should grow accordingly. This leads to a considerable computational burden. 
%
%

 In this paper, we propose a blind DL-assisted channel estimation technique for multi-user orthogonal frequency division multiplexing (OFDM) based massive MIMO systems.  
We exploit the asymptotic orthogonality of the channel vectors in very large MIMO systems to detect transmitted data symbols
without the knowledge of CIRs. In fact, if the BS antennas have spacings of larger than $\lambda /2$, where $\lambda$ is the wavelength of the carrier frequency, different users' channels will be independent, i.e., there is no antenna coupling issue or no correlation between different users' channel responses. As a result, different users' channels will be orthogonal in the asymptotic regime. Thus, by calculating the covariance matrix of the received signals at the BS antennas, we can average out the channel and noise effects. Then, we show that with some mathematical manipulation the transmitted signals can be extracted. Therefore, we detect the transmitted data symbol of each user in the channel sounding phase and consider them as virtual pilots to estimate the CIRs of all the users. However, in practice, the channel and noise effects cannot be ideally eliminated, and the remaining interference scatters the symbols and leads to an inaccurate virtual pilot estimation. To tackle this problem, we propose using denoising convolutional neural network (DnCNN) algorithm, \cite{zhang2017beyond}, to remove the remaining interference and improve the efficacy of our virtual pilot estimation, and consequently, our channel estimation technique.
Unlike other denoising algorithms, a trained DnCNN does not involve a complex optimization problem. Moreover, DnCNN model is able to perform denoising with an unknown noise level.
Finally, we assess the performance of our proposed technique through some numerical analysis. We choose the data-aided technique as a benchmark because it provides the best performance as the receiver uses known pilot sequences for channel estimation. We show that the mean square error (MSE) of our proposed technique matches that of data-aided technique. We also compare the bit error rate (BER) performance of our proposed channel estimation technique without any denoiser to the DL-assisted one in order to highlight the improvement achieved by using DnCNN. In addition, it is demonstrated that the BER curve of our proposed blind DL-assisted channel estimation technique matches the benchmark.
The rest of this paper is organized as follows. 
Section \ref{sec:SystemModel} presents the system model for the uplink of the OFDM-based multi-user massive MIMO system. 
Our proposed channel estimation technique is presented in Section \ref{sec:Proposed_Channel_Estimation}.
Then, we evaluate our proposed techniques through simulations in Section \ref{sec:NR}. Finally, the conclusions are drawn in Section \ref{sec:Conclusion}.

{\it Notations}
: Matrices, vectors and scalar quantities are denoted by boldface uppercase, boldface lowercase and normal letters, respectively.
$a[i]$ and $A[i,j]$ denote the elements $i$ and $(i,j)$ of the vector $\bm a$ and the matrix $\bm A$, respectively.
Superscripts $(\cdot)^{\rm H}$, $(\cdot)^{\rm T}$, $(\cdot)^{*}$ and $(\cdot)^{-1}$ denote Hermitian, transpose, conjugate operation and the inverse of a matrix, respectively.
$\bm I_{N}$ and $\bm 0_{N}$ are an $N \times N$ identity matrix and zero matrix, respectively. $\mathbb{E}\{\cdot \}$ denotes the expectation operator. Symbols 
$\odot$ and $\oslash$ stand for 
element-wise (Hadamard) multiplication and element-wise (Hadamard) division, respectively.
$||\bm A||$ is the Frobenius norm of the matrix $\bm A$. 
Finally, $\bm d={\rm diag}(\bm D)$ is the vector including the elements on the main diagonal of the matrix $\bm D$, and $\bm C={\rm circ}(\bm c)$ is a circulant matrix with the first column being $\bm c$.


\section{System Model}\label{sec:SystemModel}
 
Consider the uplink of an OFDM-based massive MIMO system where $P$ single antenna users are communicating with a BS equipped with $M\gg P$ antennas.
Having $N$ active subcarriers, the $N \times 1$ vector of the $\kappa^{\rm th}$ OFDM symbol for a given user $p$ before adding cyclic prefix (CP) is obtained as
\begin{equation}
\bm x_{p}^{\kappa}=\bF_{N}^{\rm H}\bm d_{p}^{\kappa},
\end{equation}
where the $N \times 1$ vector $\bm d^{\kappa}_{p}$ contains the
quadrature amplitude modulated (QAM) data symbols of user $p$ at time slot $\kappa$, which is normalized to have a power of unity. Also, $\bF_{N}$ is the normalized $N$-point DFT matrix with the elements $F_{N}[i,k]=\frac{1}{\sqrt{N}}e^{-j\frac{2\pi}{N} i k }$ for $i,k=0, \cdots, N-1$.
To absorb the channel transient response, a CP with the length larger than the length of the channel impulse response (CIR), $L$, is appended in the beginning of each OFDM symbol. 
We also consider perfect synchronization in both time and frequency.

Uplink transmission in massive MIMO systems is divided into two phases; (i) channel sounding phase in which the CIRs of all the users are estimated, and (ii) detection phase where the transmitted data symbols are detected using the estimated CIRs. 
%
As the data-aided technique, the channel sounding phase is considered to be equal to the length of $P$ OFDM symbols. Note that the data-aided technique requires to transmit number of pilot sequences equal or greater than the number of users, \cite{marzetta2006much}.
However, we assume that in the channel sounding phase, instead of transmitting known pilot sequences, users transmit their data symbols which are used as virtual pilots in section \ref{sec:Proposed_Channel_Estimation}.
In order to make the users' signals orthogonal in the channel sounding phase, we assume that users are using consecutive symbols in a round robin fashion and at each OFDM symbol only one user transmits its signal while others are silent. 
Therefore, after CP removal, the $N\times 1$ received signal vector at BS antenna $m$ for $0\leq \kappa \leq P-1 $ is given by
\begin{equation}\label{eq:rmKappaFirstP}
\bm r_{m}^{\kappa}=\bm X_{p}^{\kappa}\bm h_{m,p}+\n_{m},
\end{equation}
for $\kappa=p$, where $\bm X_{p}^{\kappa}$ is an $N\times L$ matrix including the first $L$ columns of the circulant matrix ${\rm circ}(\x_{p}^{\kappa})$,
$\n_{m} \sim \mathcal{CN}(0,\sigma^{2}_{n} {\rm \bI}_{N})$ is the complex additive white Gaussian noise (AWGN) with the variance of $\sigma^{2}_{n}$ at the $m^{\rm th}$ BS antenna. 
Note that since each user only occupies one of the first $P$ OFDM symbols in the channel sounding phase, $\bm X_{p}^{\kappa}$ is a zero matrix for $\kappa \neq p$.
In addition, $\bm h_{m,p}$ is the $L \times 1$ CIR vector between user $p$ and BS antenna $m$, and the channel taps are assumed to be a set of independent and identically distributed (i.i.d.) random variables that follow the complex normal distribution $\mathcal{CN}(0,\boldsymbol\rho_{p} )$,  where $\boldsymbol \rho_{p} $ is an $L \times 1$ vector representing the $p^{\rm th}$ user's channel power delay profile (PDP).

In the detection phase, i.e. $P \leq \kappa $, all the available time-frequency resources are being reused by all the users. Thus, the received OFDM symbol at time slot $\kappa$ after CP removal can be written as 
\begin{equation}\label{eq:rim}
\bm r_{m}^{\kappa}=\sum_{p=0}^{P-1}\bm X_{p}^{\kappa}\bm h_{m,p}+\n_{m}.
\end{equation}
After the CIRs of all the users are estimated, a linear combiner, such as matched filtering (MF), zero-forcing (ZF) or minimum mean square error (MMSE), can be used to detect transmitted data \cite{marzetta2016fundamentals}.


\section{Proposed Channel Estimation Technique}\label{sec:Proposed_Channel_Estimation}

In this section, we propose a technique to blindly estimate the data symbols transmitted at the channel sounding stage, i.e. $0\leq \kappa \leq P-1$ by exploiting the asymptotic orthogonality of the channel vectors of massive MIMO.
Then, these symbols are utilized as virtual pilots to estimate the CIR of each user at each BS antenna.

At the receiver, we take discrete Fourier transform (DFT) of the received signal in equation (\ref{eq:rmKappaFirstP}). Assuming that user $q$ has transmitted its signal on the OFDM symbol $\kappa=q$, we take the $q^{\rm th}$ received symbol to extract the $q^{\rm th}$ user's transmitted data symbols. Then, we calculate the covariance matrix of this symbol at the output of the DFT block by multiplying it to its Hermitian for $0 \leq m \leq M-1$ and average it over all the BS antennas as
\begin{align}\label{EQ:RRKappa1}
\bm R_{q} &=\frac{1}{M}\sum_{m=0}^{M-1}
\bm F_{N}
{\bm r}_{m}^{q}(\bm F_{N}{\bm r}_{m}^{q})^{\rm H}.
\end{align}
To keep the formulation simple and without loss of generality, we drop the symbol index $\kappa$ which is equal to $q$ in the rest of our derivations. Then, the equation (\ref{EQ:RRKappa1}) can be expanded as
\begin{align}\label{EQ:RRKappa}
\bm R_{q}=\bm F_{N} \bm X_{q}\bm C_{{ h_{q}},{ h_{q}}}\bm X_{q}^{\rm H}\bm F_{N}^{\rm H}+
\bm F_{N}\bm V_{q}\bm F_{N}^{\rm H}+\bm F_{N}\bm C_{{n},{n}}\bm F_{N}^{\rm H},
\end{align}
with 
\begin{align}\label{eq:ChandhEquivalent}
\bm C_{{h_{q}},{h_{q}}}&= \frac{1}{M} \sum_{m=0}^{M-1}
{\bm h}_{m,q} {\bm h}_{m,q}^{\rm H},
\end{align}
and
\begin{equation}\label{eq:VXHN}
\bm V_{q}= \bm X_{q}\bm C_{{h_{q}},{n}}
+(\bm X_{q}\bm C_{{h_{q}},{n}})^{\rm H},
\end{equation}
is the noise effect where
\begin{align}
\bm C_{{h_{q}},{n}}=\frac{1}{M}\sum_{m=0}^{M-1}{\bm h}_{m,q}{\n}_{m}^{\rm H},
\end{align}
and
\begin{align}
\bm C_{{n},{n}}=\frac{1}{M}\sum_{m=0}^{M-1}{\n}_{m}{\n}_{m}^{\rm H}.
\end{align}

According to the law of large numbers, as $M$ grows large, 
 $C_{h_{q},h_{q}}[i,j]\rightarrow \mathbb{E}\left\{ h_{m,q}[i] h_{m,q}^{*}[j] \right\}$. Since $ { h}_{m,q}[i]$ is a set of i.i.d. random variables, $\mathbb{E}\left\{h_{m,q}[i] h_{m,q}^{*}[j]  \right\}=0 $ for $ i\neq j$,
and $\mathbb{E}\left\{h_{m,q}[i] h_{m,q}^{*}[i]  \right\}=\rho_{q}[i]$ which is the $i^{th}$ element of the PDP vector of user $q$.
Thus, $\bm C_{h_{q},h_{q}}$ is a real-valued diagonal matrix with the main diagonal equal to ${\boldsymbol\rho}_{q}$. 
Note that $n_{m}[i]$ and $h_{m,q}[i]$ are also independent and both are identically distributed. Thus, as $M$ grows large, 
 $C_{h_{q},n}[i,j]\rightarrow \mathbb{E}\left\{ h_{m,q}[i] n_{m}^{*}[j] \right\}=0$, and
$C_{n,n}[i,j]\rightarrow \mathbb{E}\left\{ n_{m}[i] n_{m}^{*}[j] \right\}$ which is zero for $i \neq j$ and $\sigma_{n}^{2}$ for $i=j$ for all $i$ and $j$. Thus, the last term on the right hand side of equation (\ref{EQ:RRKappa}) can be removed by subtracting $\sigma_{n}^{2}\bm I_{N}$.
Then, considering $\bm F_{N}^{\rm H} \bm F_{N}=\bm I_{N}$, equation (\ref{EQ:RRKappa}) can be rewritten as
\begin{align}
\bm R_{q} &= (\bm F_{N}\bm X_{q}\bm F_{N}^{\rm H})(\bm F_{N}\bm C_{ h_{q},h_{q}}\bm F_{N}^{\rm H})(\bm F_{N}\bm X_{q}^{\rm H}\bm F_{N}^{\rm H})+
\bm F_{N}\bm V_{q}\bm F_{N}^{\rm H}.
\end{align}
It is worth noting that since $\bm X_{q}$ is a circulant matrix, $\bm F_{N}\bm X_{q}\bm F_{N}^{\rm H}$ is a diagonal matrix with the main diagonal equal to $\bm d_{q}$. Hence, we have
\begin{align}
\bm R_{q} &=\rm diag(\bm d_{q})\widetilde{\bm C}_{h_{q},h_{q}}
\rm diag(\bm d_{q}^{\rm H})+
\bm F_{N}\bm V_{q}\bm F_{N}^{\rm H}\nonumber \\ 
&=\widetilde{\bm C}_{h_{q},h_{q}} \odot \left\{
\bm d_{q}\bm d_{q}^{\rm H}
\right\}
+
\bm F_{N}\bm V_{q}\bm F_{N}^{\rm H},
\end{align}
where $\widetilde{\bm C}_{h_{q},h_{q}}=\bm F_{N}\bm C_{ h_{q},h_{q}}\bm F_{N}^{\rm H}= \rm circ(\bar{\boldsymbol \rho}_{q})$ is a circulant matrix as $\bm C_{ h_{q},h_{q}}$ is diagonal, and $\bar{\boldsymbol \rho}_{q}$ is $N$-point DFT of $\boldsymbol \rho_{q}$. Furthermore, if we element-wise divide the matrix $\bm R_{q}$ by the elements of the matrix $\widetilde{\bm C}_{h_{q},h_{q}} $ as
\begin{align}
 \bm G_{q}&=\widetilde{\bm C}_{h_{q},h_{q}} \oslash \bm R_{q}
\nonumber \\ 
&=
\bm d_{q}\bm d_{q}^{\rm H}+\widetilde{\bm V}_{q},
\end{align}
where 
\begin{align}
\widetilde{\bm V}_{q}=\left\{\widetilde{\bm C}_{h_{q},h_{q}}\oslash(\bm F_{N}\bm V_{q}\bm F_{N}^{\rm H}) 
\right\}.
\end{align}

Let us assume that the data on the first subcarrier of the first OFDM symbols of the users, i.e. $d_{p}[0]$ for $p=0,1,\cdots,P-1$, are known as reference symbols. Then, we can form a matrix $\bm Y_{q}$ as
\begin{align}\label{eq:chap_est_Ydatamatrix}
 Y_{q}[i,j]&=  d_{q}[0]\times \frac{  G_{q}[i,j]}{G_{q}[0,j]}  \nonumber \\
&= d_{q}[0]\times\frac{ d_{q}[i] d_{q}[j]+\widetilde{V}_{q}[i,j]}{ d_{q}[0]d_{q}[j]+\widetilde{V}_{q}[0,j]} 
 \nonumber \\
&= w_{q}[j]d_{q}[i]
+\widehat{V}_{q}[i,j],
\end{align}
where
\begin{align}
w_{q}[j]=\frac{ d_{q}[0] d_{q}[j]}{ d_{q}[0]d_{q}[j]+\widetilde{V}_{q}[0,j]},
\end{align}
and
\begin{align}
\widehat{V}_{q}[i,j]=\frac{ d_{q}[0]\widetilde{V}_{q}[i,j]}{ d_{q}[0]d_{q}[j]+\widetilde{V}_{q}[0,j]}.
\end{align}


One can see from equation (\ref{eq:chap_est_Ydatamatrix}) that all the $N$ elements in the $i^{\rm th}$ row of the matrix $\bm Y_{q}$ contain the information of $d_{q}[i]$. Moreover, as the number of BS antennas, $M$, grows large, noise effect is averaged out and $V_{q}[i,j]$ in equation (\ref{eq:VXHN}) tends zero. Thus, we have $w_{q}[j]\approx 1$ and $\widehat{V}_{q}[i,j]\approx 0$. Therefore,
the data symbols $d_{q}[i]$ can be estimated as
\begin{align}\label{eq:relaYandD}
\hat{d}_{q}[i]=\frac{1}{N_{\rm a}}\sum_{j=0}^{N_{\rm a}-1}Y_{q}[i,j],
\end{align}
for $i=0,1, \cdots, N-1$. However, due to the limited number of BS antennas in practical systems, noise effects will not completely fade away and distort the received symbols. 


To address this issue, we propose applying CNN-based DL algorithms called DnCNN to mitigate noise effect and extract the information in virtual pilots with a higher level of accuracy. 
DnCNN contains three types of layers. The first convolutional layer uses $64$ different $3\times 3\times 1$ filters and is followed by a rectified linear unit (ReLU). Second, there are $18$ convolutional layers with $64$ filters of size $3\times 3\times 64$, batch-normalization, \cite{ioffe2015batch}, and a ReLU. Final layer utilizes one $3\times 3\times 64$ filter to reconstruct the output. 
DnCNN is capable to learn either the original mapping or the residual mapping. In original mapping, DnCNN directly detects the data from a noisy observation. While in residual mapping, first, it detects additive noise, and then, subtracts it from the input to recover the original data. This method is known as residual learning, \cite{he2016deep}. 
Here, we deploy DnCNN for the residual mapping.  
As DnCNN only takes real-valued input, we separate the real and the imaginary parts of the matrix $\bm Y_{q}$ into two $N\times N$ matrices and give them to DnCNN as two different inputs.
Then, we can calculate the complex noise from the outputs of DnCNN corresponding to the real and imaginary parts of the noise as
\begin{align}
 \widehat{\bm V}^{(\rm DnCNN)}_{q} = \widehat{\bm V}^{(\rm DnCNN)}_{q,{\rm real}} + \widehat{\bm V}^{(\rm DnCNN)}_{q,{\rm imag}}.
\end{align}
By subtracting the noise from $\bm Y_{q}$, we have the denoised version of the matrix $\bm Y_{q}$ as
\begin{align}
 Y^{(\rm DnCNN)}_{q}[i,j] = Y_{q}[i,j]-\widehat{V}^{(\rm DnCNN)}_{q}[i,j],
\end{align}
and the data symbols $d_{q}[i]$ can be estimated as
\begin{align}
\hat{d}_{q}^{(\rm DnCNN)}[i]=\frac{1}{N}\sum_{j=0}^{N-1}Y_{q}^{(\rm DnCNN)}[i,j],
\end{align}
for $i=0,1, \cdots, N-1$. These symbols are then deployed as virtual pilots to extract the channel estimates of all the users at the BS.

%
Thus, after detecting the users' virtual pilots, we employ a simple maximum likelihood (ML) channel estimation to estimate the channel impulse responses. To this end, we rearrange equation (\ref{eq:rim}) as
\begin{equation}
\bm r_{m}=\bm X\bm h_{m}+\n_{m},
\end{equation}
where $\bm X=[\bm X_{0},\bm X_{1},\cdots,\bm X_{P-1}]$, and $\bm h_{m}=[\bm h_{m,0}^{\rm T},\bm h_{m,1}^{\rm T}, \cdots,\bm h_{m,P-1}^{\rm T}]^{\rm T}$. Then, the logarithm of the conditional probability density function (PDF) of $\bm r_{m}$ given $\bm h_{m,p}$ can be written as
\begin{align}\label{eq:chanest}
\ln {\it p}(\bm r_{m}|\bm h_{m,p}) &= \Theta_{0}-\Theta_{1}\left[\bm r_{m}-\bm X \bm h_{m,p}\right]^{\rm H} \\ \nonumber &\times \left[\bm r_{m}-\bm X \bm h_{m,p} \right],
\end{align}
where $\Theta_{0}=-N \ln (2 \pi \sigma^{2}_{n})/2$, and $\Theta_{1}=1/2\sigma^{2}_{n}$. By taking the derivative of equation (\ref{eq:chanest}) with respect to $\bm h_{m,p}$ and setting it to zero, the estimated channels of all the users at a given BS antenna $m$ can be obtained as

\begin{equation}\label{eq:chanEstimated}
 \hat{\bm h}_{m}=(\bm X^{\rm H}\bm X)^{-1}\bm X^{\rm H} \bm r_{m}.
\end{equation}

%


\section{Numerical Analysis}\label{sec:NR}

\begin{figure}
		\includegraphics[scale=0.65]{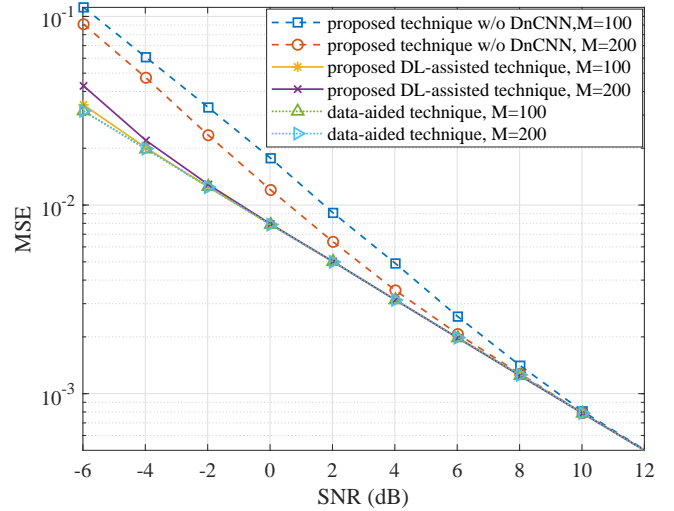}
		\caption{MSE with respect to SNR for $P=4$ users, $N = 128$ subcarriers, and $16$-QAM modulation, and different number of BS antennas.}
\label{fig:MSE_16QAM_4U}
\end{figure}

\begin{figure}
		\includegraphics[scale=0.65]{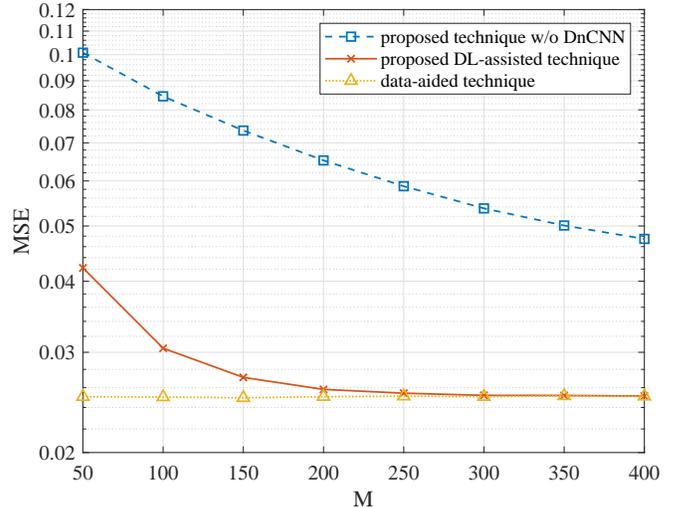}
		\caption{MSE with respect to the number of BS antennas, M,  for $P=4$ users and $N = 128$ subcarriers with $16$-QAM modulation at SNR=$-5$ dB.}
\label{fig:MSE_M_32_all}
\end{figure}

In this section, we evaluate our theoretical developments in the previous section through numerical analysis. We consider a massive MIMO setup with $P=4$ users which are transmitting 16-QAM symbols over $N=128$ subcarriers. In our simulation, we use the 3GPP Long Term Evolution (LTE) extended typical urban (ETU) channel model, \cite{yang2014lte}, and the CP length of $N/8$.
We assume perfect power control for all the users. 
Using similar numerology as in LTE, channel coherence time is set to $1$ ms.
We also consider the worst case where the coherence frequency is equal to the subcarrier spacing.
The training and validation sets for training DnCNN contains 80000 and 16000 samples. Moreover, the data is scaled to the range of $[0,1]$ to be applied to DnCNN. Adam optimizer, \cite{kingma2014adam}, is utilized to train the networks. 
As a benchmark, we consider a data-aided channel estimation technique in which known pilot symbols are transmitted at the channel sounding stage \cite{bjornson2015massive}. Note that the data-aided channel estimation technique provides the best estimation of the channels since the pilot symbols are known to the receiver. 

Fig~\ref{fig:MSE_16QAM_4U} shows the MSE performance of our proposed blind channel estimation technique for different number of BS antennas as a function of signal to noise ratio (SNR).
As it is mentioned in section \ref{sec:Proposed_Channel_Estimation}, the remaining noise in equation (\ref{eq:chap_est_Ydatamatrix})
leads to inaccurate
virtual pilot detection. This causes a considerable gap between the performance of the proposed technique without DnCNN and the benchmark.
As it is demonstrated, this gap is removed by applying DnCNN denoiser. It is shown that the MSE of our proposed DL-assisted technique is very close to the benchmark by deploying $M=100$ BS antennas.
Moreover, by increasing the number of BS antennas, the MSE gets to the benchmark even at low SNRs. 

\begin{figure}
		\includegraphics[scale=0.65]{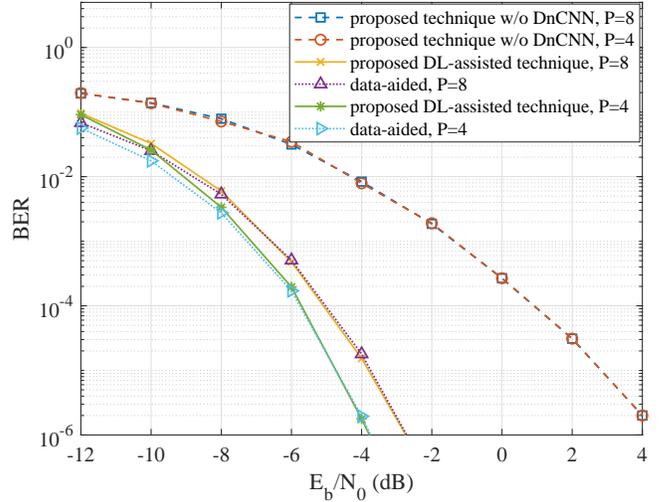}
		\caption{BER for $M=100$ BS antennas, $N = 128$ subcarriers, $16$-QAM modulation, and different number of users.}
\label{fig:BERall_ZF_32}
\end{figure}

\begin{figure}
		\includegraphics[scale=0.65]{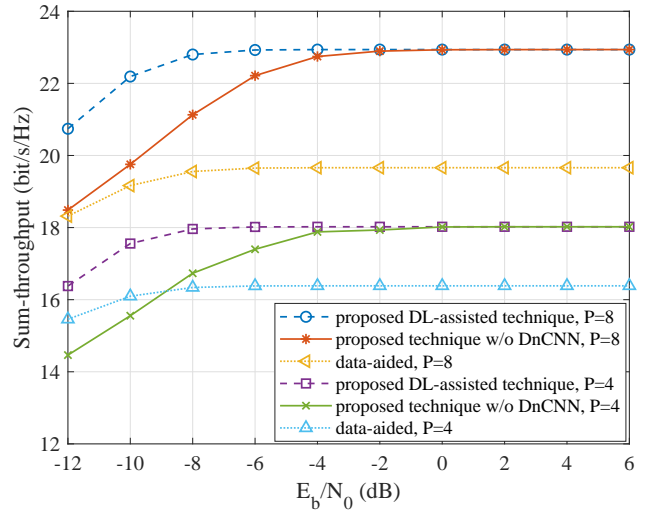}
		\caption{Throughput for $M=100$ BS antennas, $N = 128$ subcarriers, and $16$-QAM modulation.}
\label{fig:ThPut_MF_32}
\end{figure}

In Fig~\ref{fig:MSE_M_32_all}, we demonstrate the effect of the number of BS antennas on the MSE performance of our proposed technique at SNR$=-5$ dB. As it is shown, increasing the number of BS antennas can reduce the MSE of our proposed technique without DnCNN, however, it still has a noticeable distance from the benchmark. This is while the MSE of our DL-assisted technique reaches the benchmark by using $M=200$ BS antennas.  
%
 %
In Fig~\ref{fig:BERall_ZF_32}, we compare the BER performance of our proposed virtual pilot detection technique without DnCNN and the one assisted by DnCNN to the BER of the data-aided technique for different number of users.
We illustrate the improvement achieved by the DL-based denoiser in our proposed technique. In addition, it is shown that the BER performance of our DL-assisted virtual pilot and channel estimation technique matches that of the data-aided channel estimation in both cases of $P=4$ and $P=8$ users. 
Furthermore, we plot the sum-throughput of our proposed technique with respect to $\rm E_{b}/N_{0}$ in Fig~\ref{fig:ThPut_MF_32}. This figure shows the superior performance of
our proposed technique compared to the data-aided technique.

\section{Conclusion}\label{sec:Conclusion}

In this paper, we proposed a blind DL-assisted channel estimation technique for multi-user OFDM-based massive MIMO systems. 
We introduced a novel approach to extract transmitted data symbols without the knowledge of CIRs. Then, treating the detected symbols as virtual pilots, we estimated the users' CIRs. 
To make our solution applicable to practical systems with a limited number of BS antennas, we deployed a DL-assisted denoising algorithm called DnCNN. A DnCNN algorithm can be trained to perform denoising with an unknown noise level. In addition, it does not contain any optimization problem when it is used on-line.
DnCNN mitigates the remaining interference of the channels and noise, and it considerably improves the performance of our proposed virtual pilot detection, and consequently, our channel estimation technique.
To evaluate the performance of our proposed solution,
we compared our blind channel estimation technique to the data-aided technique as a benchmark since it is based on known pilot signals and provides the best channel estimation performance.
We demonstrated that there is a gap between the BER performance of our virtual pilot estimation technique without the denoising algorithm and the benchmark, which is removed by applying DnCNN. 
Moreover, we showed that the MSE of our proposed DL-assisted technique matches that of the data-aided channel estimation technique. We also demonstrate the superiority of our proposed DL-assisted technique compared to the data-aided case in terms of sum-throughput.

%
\bibliographystyle{IEEEtran}

\end{document}